\pdfoutput=1
\documentclass[a4paper]{llncs}
\usepackage[utf8]{inputenc}
\usepackage{graphicx}
\usepackage{amsmath}
\newcommand{\ie}{i.\,e. }
\newcommand{\eg}{e.\,g. }
\newcommand{\cf}{cf. }
\newcommand\blfootnote[1]{%
  \begingroup
  \renewcommand\thefootnote{}\footnote{#1}%
  \addtocounter{footnote}{-1}%
  \endgroup
}
\begin{document}
\title{Design, Analysis and Evaluation of Control Algorithms for Applications in Smart Grids}
%
\author{Christian Hinrichs \and Michael Sonnenschein}
%
\institute{%
  Department of Computing Science, University of Oldenburg, Germany,\\
  \email{christian.hinrichs@uni-oldenburg.de, sonnenschein@informatik.uni-oldenburg.de}
}
\maketitle
\begin{abstract}
  In many countries, the currently observable transformation of the power supply system from a centrally controlled system towards a complex ``system of systems'', comprising lots of autonomously interacting components, leads to a significant amount of research regarding novel control concepts. To facilitate the structured development of such approaches regarding the criticality of the targeted system, the research and development of a distributed control concept is demonstrated by employing an integrated methodology comprising both the Smart Grids Architecture Model framework (SGAM) and the Smart Grid Algorithm Engineering process model (SGAE). Along the way, a taxonomy of evaluation criteria and evaluation methods for such approaches is presented. For the whole paper, the Dynamic Virtual Power Plants business case (DVPP) serves as motivating example.
  \keywords{Heuristics $\cdot$ Distributed Problem Solving $\cdot$ Self-Organization $\cdot$ Synchronous vs. Asynchronous Approaches $\cdot$ Virtual Power Plants}
\end{abstract}
\blfootnote{To be published in: Marx Gómez, Michael Sonnenschein et al. (Eds.), Advances and new Trends in Environmental Informatics: Selected and Extended Contributions from the 28th International Conference on Informatics for Environmental Protection, Springer}
\enlargethispage{15mm}
\section{Introduction}\label{sec:introduction}
A significant share of global CO$_2$ emissions can be explained by the combustion of fossil fuels for power production. Hence, it has become politically widely accepted in Europe, to reduce national shares of fossil fuels in power production significantly. Such a politically driven evolution of the power system faces not only economical and societal challenges, but it must also address several technological challenges of ensuring a highly reliable power supply, as described in e.g. \cite{Appelrath2012}. In order to address
these challenges, new concepts for power grid operation are needed. The notion of Smart Grids has been introduced for this purpose. The European Technology Platform for Electricity Networks of the Future defines a Smart Grid as an ``electricity network that can intelligently integrate the actions of all users connected to it -- generators, consumers and those that do both -- in order to efficiently deliver sustainable, economic and secure electricity supplies.'' \cite{ETP2010} However, this implies an increased computational complexity for optimizing the coordination of these individually configured, distributed actors. A significant body of research currently concentrates on this topic, see \eg the research agenda proposed in \cite{Ramchurn2012}. In turn, new possibilities are opened up for business players to offer novel control concepts also to distributed generators and consumers.
\par
\begin{figure}
  \centering
  \includegraphics[width=0.97\textwidth]{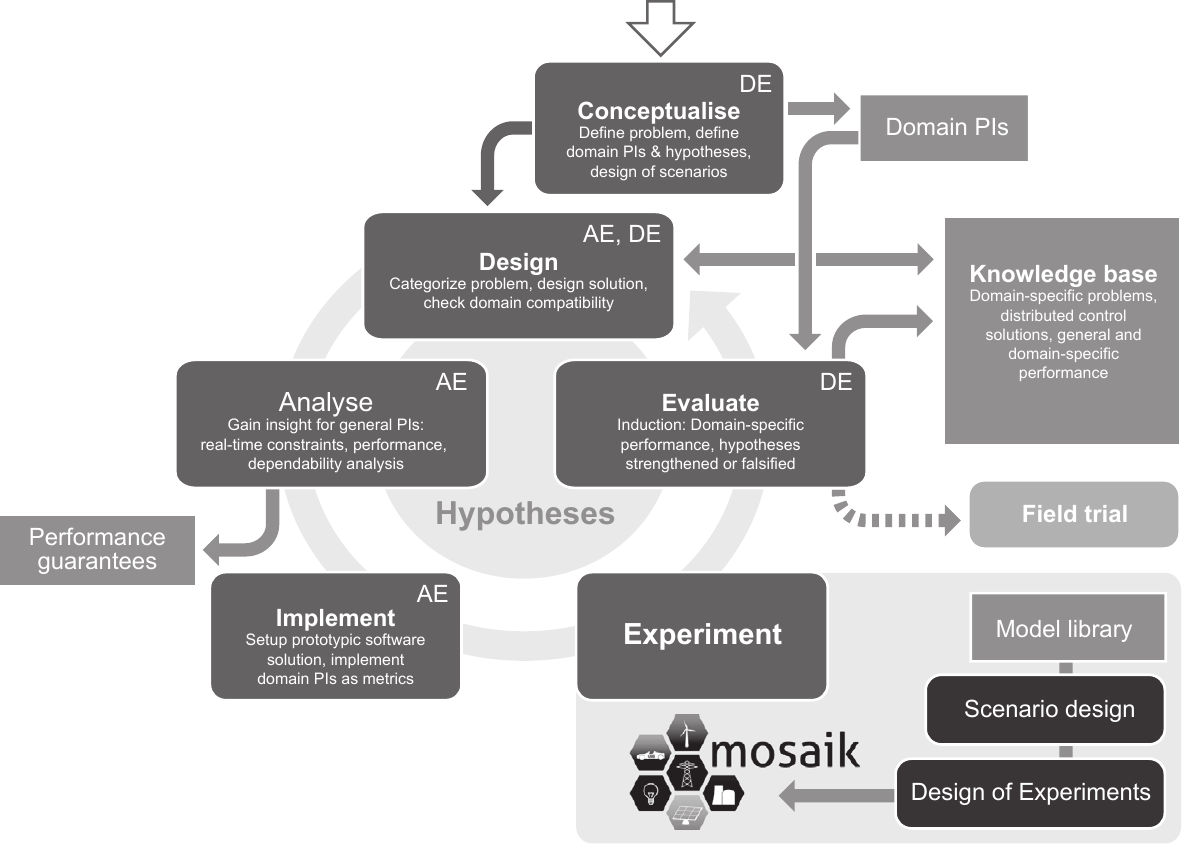}
  \caption{The Smart Grid Algorithm Engineering process model (SGAE), taken from \cite{Niesse2014} (reproduced with permission).}
  \label{fig:SGAE}
\end{figure}

 The power supply system is a critical infrastructure, therefore such approaches must be carefully studied in a secure environment before being implemented in the field. For gaining reliable results, however, this secure environment should reflect as many significant properties as possible of the targeted application area. Two relevant methodologies have been proposed to support the development of Smart Grid applications in this sense:
\begin{enumerate}
  \item The Smart Grids Architecture Model framework (SGAM) provides a way to document static overviews of systems and actors in a Smart Grid use case \cite{CEN2012}. As it lacks a dynamic view as well as the annotation of (non-)functional requirements for interfaces, it is complemented by the use case template IEC 62559 \cite{Trefke2013}.
  \item The Smart Grid Algorithm Engineering process model (SGAE) introduces guidelines for application-oriented research and development especially in control algorithms for power systems \cite{Niesse2014}, see Fig.\,\ref{fig:SGAE}.
\end{enumerate}
While the SGAM focuses on the software structural design and realization of an intended system behavior (\cf\,\cite{Santodomingo2014}), the SGAE understands itself as an engineering approach to develop validated algorithms for this specific domain. But as the latter is a cyclic process model that runs through the phases \emph{Design}, \emph{Analyze}, \emph{Implement}, \emph{Experiment} and \emph{Evaluate} iteratively, the SGAM can easily be integrated in different stages of the SGAE (some of which have already been proposed in \cite{Niesse2014}).
\par In the contribution at hand, we demonstrate the utilization of this integrated process model for the development of a distributed control concept for the Smart Grid. Reflecting a full cycle of the SGAE, the remainder of this paper is structured as follows:
\begin{itemize}
  \item The motivating business case ``Trading Flexibility in the Smart Grid'' along with a principal control concept is described in Sect.\,\ref{sec:conceptualize}. This corresponds to the initial \emph{Conceptualize} phase in the SGAE.
  \item For a specific use case in the motivating business case, the \emph{Design} of a concrete algorithm as well as a suitable architecture in compliance with international standards is demonstrated by employing the SGAM framework in Sect.\,\ref{sec:design}. This allows for checking the compatibility of the use case and the included control concept to the Smart Grid domain.
  \item In Sect.\,\ref{sec:analysis}, the \emph{Analysis} and \emph{Evaluation} of the control concept is considered from a more general perspective. While these phases are usually treated separately, we take a combined approach here in order to present a novel taxonomy of evaluation criteria.
  \item A \emph{Dependability Analysis} of the designed algorithm is undertaken in order to derive performance guarantees. This is done by deriving intrinsic properties of the approach formally. A brief overview on this part is given in Sect.\,\ref{sec:analyze}.
  \item Section\,\ref{sec:evaluation} discusses empirical methods for the evaluation of Smart Grid control concepts. Finally, an example of the control algorithm developed in this contribution is given.
\end{itemize}
As this contribution is an extended and revised version of \cite{HS14a}, Sect.\,\ref{sec:analysis} and Sect.\,\ref{sec:evaluation} reflect the main concepts from \cite{HS14a}, while the remaining parts present novel content.
\section{Conceptualize: Trading Flexibility as Business Case}\label{sec:conceptualize}
The traditional power supply system can be seen as a centralized system. It consists of only a small number of controllable power plants. A control center acts as a central component that knows the operational constraints of the plants and performs a scheduling of the plants' operations with respect to demand and weather forecasts as well as the grid status and possibly the market situation. However, as already indicated in the introduction, such a control paradigm is not suitable for future Smart Grids anymore. It is widely accepted that the power supply system of the future will be characterized by a distributed architecture comprising autonomous components with individual sub-objectives, see \eg \cite{Wu2005,Ilic2007,McArthur2007,Uslar2013}. In order to orchestrate those components towards global stability and reliability of the system, appropriate control mechanisms are necessary.
\par With the introduction of the \emph{Flexibility Concept} (and therein the notions of \emph{Flexibility Providers} and \emph{Flexibility Operators}), the CEN-CENELEC-ETSI Smart Grid Coordination Group (SG-CG) depicts a possible architecture for such a distributed architecture \cite{CEN2012a}. A Flexibility Provider is described as an entity offering flexibility in generation, load or storage of electrical power. In contrast to the traditional power supply system, these entities can be of very small scale (\eg individual households or appliances). Following, they do not participate directly in energy markets, but are contracted with Flexibility Operators instead, which aggregate the flexibilities of many units and make use of them in the grid or in energy markets. Hence, the Flexibility Operators act as a coordination layer between the grid/market on the one hand, and the Flexibility Providers on the other hand. From a business perspective, Flexibility Operators can be seen as Energy Service Providers (ESP), offering services that enable the customers to trade their aggregated flexibilities in the market. Of course, besides providing the required technical infrastructure, the Flexibility Operators have to employ sophisticated business logic for this task, \ie suitable aggregation and optimization methods.
\par
\begin{figure}
  \centering
  \includegraphics[width=\textwidth]{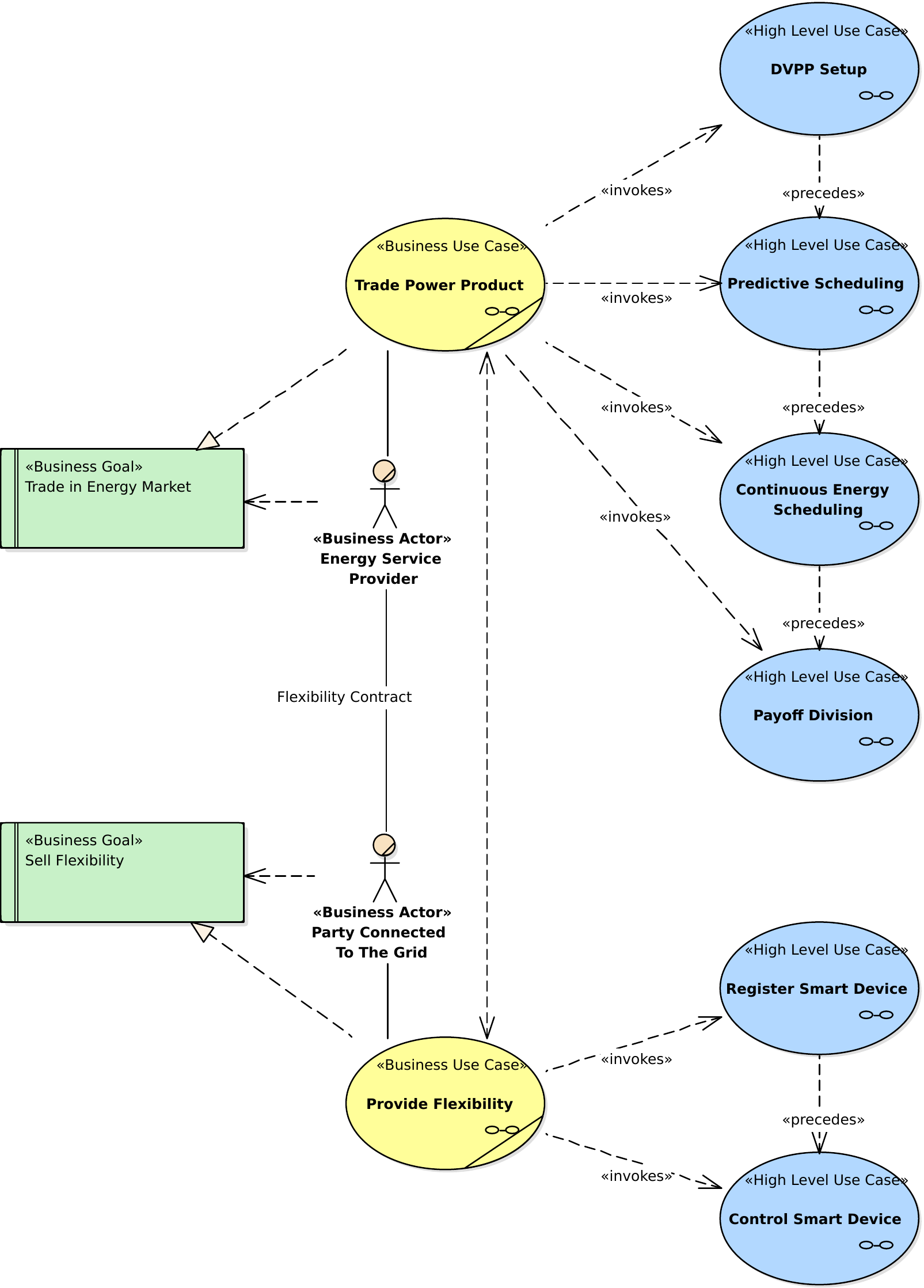}
  \caption{Business case overview of the DVPP concept.}
  \label{fig:business-case}
\end{figure}

 As an example of such a business logic, we refer to the concept of Dynamic Virtual Power Plants (DVPPs), which was initially introduced in \cite{Niesse2012} and has been reformulated in \cite{SHNV15}. The concept is characterized by aggregating decentralized power producers, local storage systems and controllable loads in a self-organized way with respect to concrete products in an energy market. This way, multiple coalitions of Flexibility Providers are formed, where each coalition offers an individual power product to the market. After delivering a product, a coalition dissolves and the former participating units can then self-determinedly join the formation process of other coalitions for subsequent tradable energy products. In particular, the DVPP concept comprises the following subprocesses:
\begin{enumerate}
  \item\label{step:dvpp-setup}\emph{DVPP Setup:} Flexibility Providers are aggregated to DVPPs by coalition formation, such that the members of each DVPP agree upon trading a specific power product in the market (\eg a certain block product in an electricity spot market). Bids for these products are then placed in the market.
  \item\label{step:predictive-scheduling}\emph{Predictive scheduling:} After a successful bid, a DVPP is obliged to deliver the power product. For this, the members of the DVPP have to be scheduled within their individually defined flexibility. This is done prior to the actual delivery of the product in a predictive scheduling process.
  \item\label{step:continuous-scheduling}\emph{Continuous scheduling:} To compensate for unforeseen changes or forecast errors, a continuous scheduling is performed during the delivery of the product. Here, the unit's schedules are adapted such that product delivery is not endangered.
  \item\label{step:payoff-division}\emph{Payoff division:}  Subsequently, the revenues gained from product delivery are distributed among the DVPP members.
\end{enumerate}
Referring to the Flexibility Concept of the SG-CG above, an ESP in the DVPP concept provides the technical infrastructure as well as management functions for the formation and operation of DVPPs, and thus realizes the role of a Flexibility Operator. A possible model of this business case is shown in Fig.\,\ref{fig:business-case} using the Unified Modeling Language (UML) notation.\footnote{\url{http://www.omg.org/spec/UML}}
Please note that we restrict our model to commercial DVPPs only and thus neglect the possibility to form technical DVPPs for providing ancillary services as described in \cite{Niesse2012}. The diagram shows both the Flexibility Operator as ESP and the Flexibility Provider as customer (which corresponds to a Party Connected to the Grid in the ENTSO-E harmonized electricity market role model, \cite{ENTSO-E2014}) with their associated business goals and the according business use cases that realize the goals. Attached to the business use cases are several high level use cases (HLUC), which serve as placeholders for the underlying processes that constitute the respective business use cases. On the customer side, only the initial registration of the Smart Device (\ie a decentralized power producer, local storage system or controllable load) with the ESP, and the actual control of this device according to received schedules are located, as these have to be undertaken by the customer himself. All other HLUC, corresponding to the individual subprocesses of the DVPP concept outlined above, are attributed to the ESP.
\par In this form, the business case and its control concept give a rough idea of the intended system behavior. In the following section, an architectural design will be developed for this system. In particular, the decomposition of a HLUC into several primary use cases (PUC) and their precise localization in an architecture based on standards is shown using the SGAM framework. For demonstration purposes, we focus on step\,\ref{step:predictive-scheduling} of the DVPP concept, \ie the HLUC \emph{Predictive Scheduling}.
\section{Design: Predicitive Scheduling by Distributed Control}\label{sec:design}
As motivated in \cite{Niesse2012}, the DVPP concept employs self-organization mechanisms for each of its subprocesses. Thus the HLUC \emph{Predictive Scheduling} is realized with an asynchronous distributed heuristic in the following (more information about this type of heuristics can be found in \cite{Lynch1996,HS14b}). The main aspect here is that each participating DVPP member (\ie each Smart Device as Flexibility Provider) acts on its own behalf during the scheduling process, but in such a way that a global optimization towards a satisfying schedule assignment for all DVPP members emerges. We will recap the approach briefly.
\par First of all, the HLUC \emph{Predictive Scheduling} requires that the HLUC \emph{DVPP Setup} has finished (\cf Fig.\,\ref{fig:business-case}), so that one or more DVPP have formed with respect to concrete power products. Then, for a given DVPP, the associated power product constitutes the optimization target for the scheduling heuristic. For the sake of convenience, we model the power product as a power profile that has to be realized by the DVPP. Now the task is to find a schedule assignment for each participating Smart Device over the planning horizon specified by the power product, such that the aggregation of all selected schedules matches the target power profile as close as possible. When each Smart Device then executes its assigned schedule later on, the DVPP effectively delivers the agreed power product.
\par
\begin{figure}
  \centering
  \includegraphics[width=0.97\textwidth]{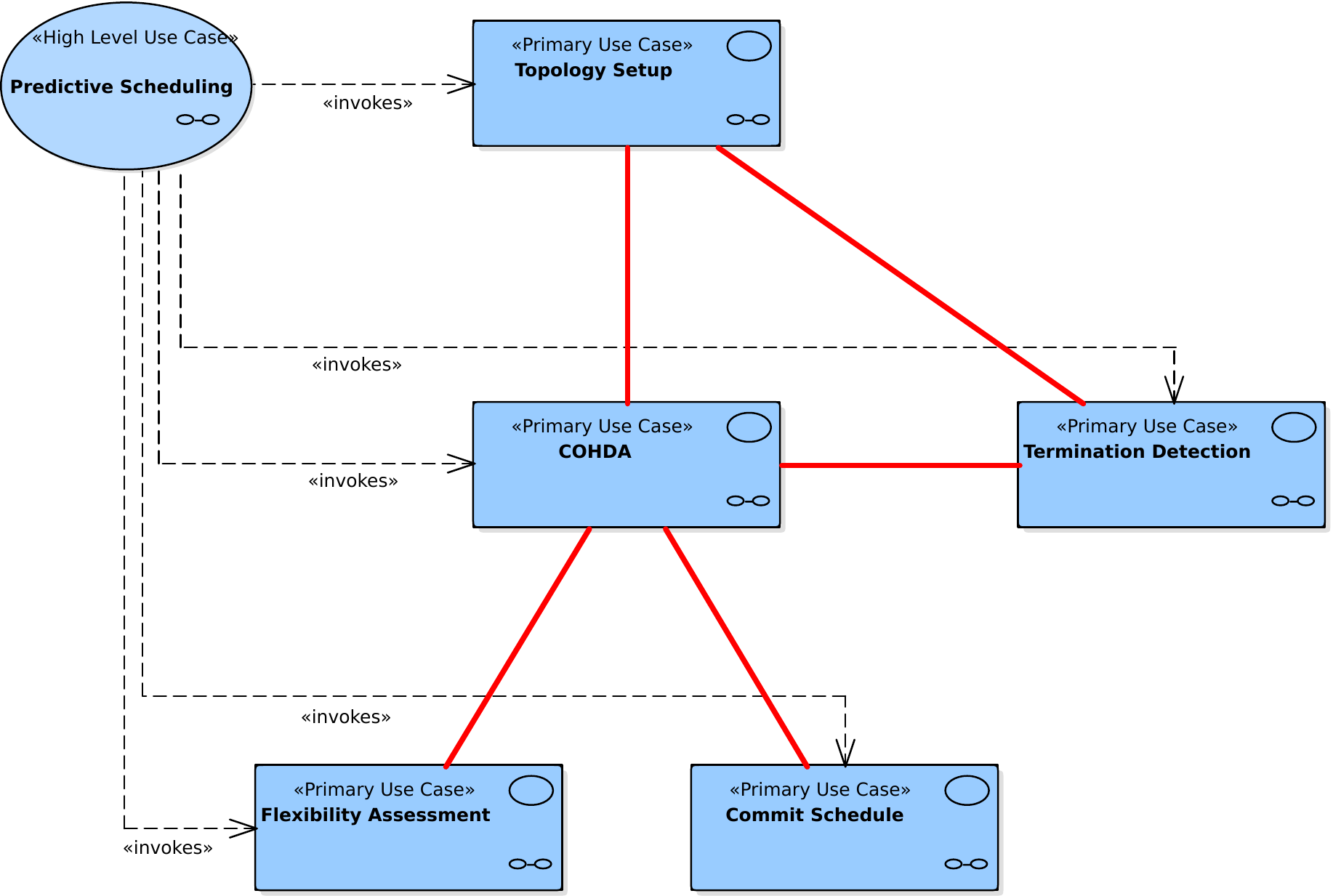}
  \caption{Decomposition of the HLUC \emph{Predictive Scheduling} into several PUCs.}
  \label{fig:hluc-predictive-scheduling}
\end{figure}

 This optimization task can be formulated as a distributed combinatorial optimization problem, for which a suitable heuristic has been proposed in \cite{HLS14a}: the Combinatorial Optimization Heuristic for Distributed Agents (COHDA). Operating on a virtual communication topology, the heuristic performs the scheduling task in a fully distributed way. This is done by letting the individual Flexibility Providers coordinate based on a few simple behavioral rules for each participant, which are triggered by message exchanges between them. In Fig.\,\ref{fig:hluc-predictive-scheduling}, the execution of the heuristic is shown in the context of the HLUC \emph{Predictive Scheduling}.
Here, the PUC COHDA is surrounded by other, supportive PUCs. All PUCs are invoked by the superordinate HLUC, while the thick lines denote functional interrelations. More precisely, as COHDA relies on a virtual communication topology, the antecedent PUC \emph{Topology Setup} has to be carried out, such that a suitable communication topology is created for the respective DVPP.
\par On the other hand, each participating DVPP member, \ie each Flexibility Provider, needs a collection of feasible schedules to choose from during the scheduling process. Hence, also the PUC \emph{Flexibility Assessment} is carried out prior to COHDA, in which a Smart Device determines its own flexibility for the planning horizon. Because the heuristic operates in a fully distributed way, the PUC \emph{Termination Detection} might be invoked at arbitrary points in time during the operation of COHDA in order to check for (and eventually announce) the termination of the heuristic. After termination, the resulting schedules are committed to the respective Smart Devices (PUC \emph{Commit Schedule}).
\par
\begin{figure}
  \centering
  \includegraphics[width=0.97\textwidth]{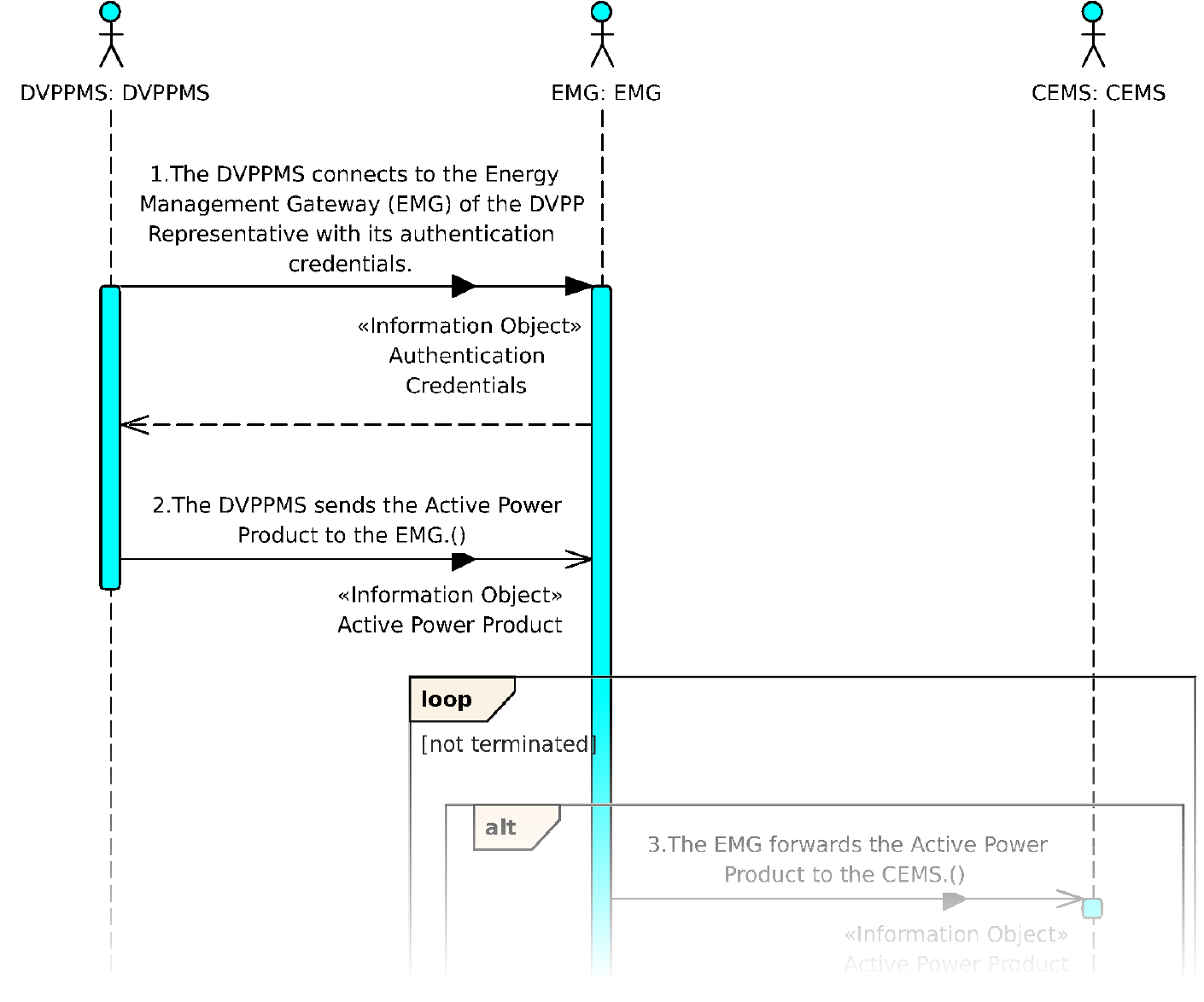}
  \caption{Beginning of a sequence diagram for the COHDA scheduling approach.}
  \label{fig:puc-cohda}
\end{figure}

 While this abstraction layer allows modeling of the interrelations of the PUCs, the concrete localization of the business logic to actual devices needs further refinement. Thus, for each PUC, the participating actors and their information exchange is identified using scenario specifications. For the PUC \emph{COHDA}, such a specification is visualized with the beginning of an according sequence diagram in Fig.\,\ref{fig:puc-cohda}.
Therein logical actors have been tailored from related generic use cases (\cf \cite{CEN2012a}) for the HLUC \emph{Predictive Scheduling}:
\begin{itemize}
  \item The DVPP Management System (DVPPMS) represents the ESP on a technical layer and constitutes a system that performs management operations regarding the complete lifecycle of DVPPs. This system also serves as a connection point to the energy market and thus embodies the role of a Flexibility Operator.
  \item The Energy Management Gateway (EMG) is an access point in the customer premises that communicates between external and internal systems.
  \item The Customer Energy Management System (CEMS) is responsible for gathering flexibilities in the customer premises as well as for performing optimization tasks regarding flexibility contracts.
\end{itemize}
The information flow between these actors is depicted in the figure using the standardized UML notation. An important aspect here is the attachment of \emph{Information Objects} to the information flows, allowing modeling of the concrete data that has to be transferred between the actors.
\par More details on the depicted coordination steps can be found in \cite{HLS14a,HLS14b}.
\subsection{The SGAM Layers}\label{sec:sgam-layers}
With the modelling of the surrounding business case, its associated HLUCs, the involved PUCs as well as a detailed specification of the PUCs internal operations, resulting in the identification of involved actors and information objects, one may now begin to physically lay out the system in an architecture based on standards. The SGAM methodology defines five interoperability layers for this task \cite{Niesse2014}:
\begin{itemize}
  \item The \emph{Business Layer} presents a view to model interrelations regarding business, regulatory and market aspects.
  \item The \emph{Function Layer} reflects the interrelations between functions and services according to use cases from a surrounding business case.
  \item The \emph{Information Layer} formally describes the exchanged data between functions/components in terms of standardized information objects.
  \item The \emph{Communication Layer} defines concrete protocols and mechanisms for the data exchange between physical components according to the identified information flows.
  \item The \emph{Component Layer} maps logical actors of the above layers to physical components in the Smart Grid context.
\end{itemize}
Each abstraction layer spans the \emph{Smart Grid Plane}, which allows to localize entities with respect to both electrical process and information management viewpoints. The former viewpoint is subdivided into several physical domains (\eg Generation, Transmission, Distribution \dots) while the latter viewpoint comprises a number of hierarchical zones (\eg Market, Enterprise, Operation \dots).
\par
\begin{figure}
  \centering
  \includegraphics[width=\textwidth]{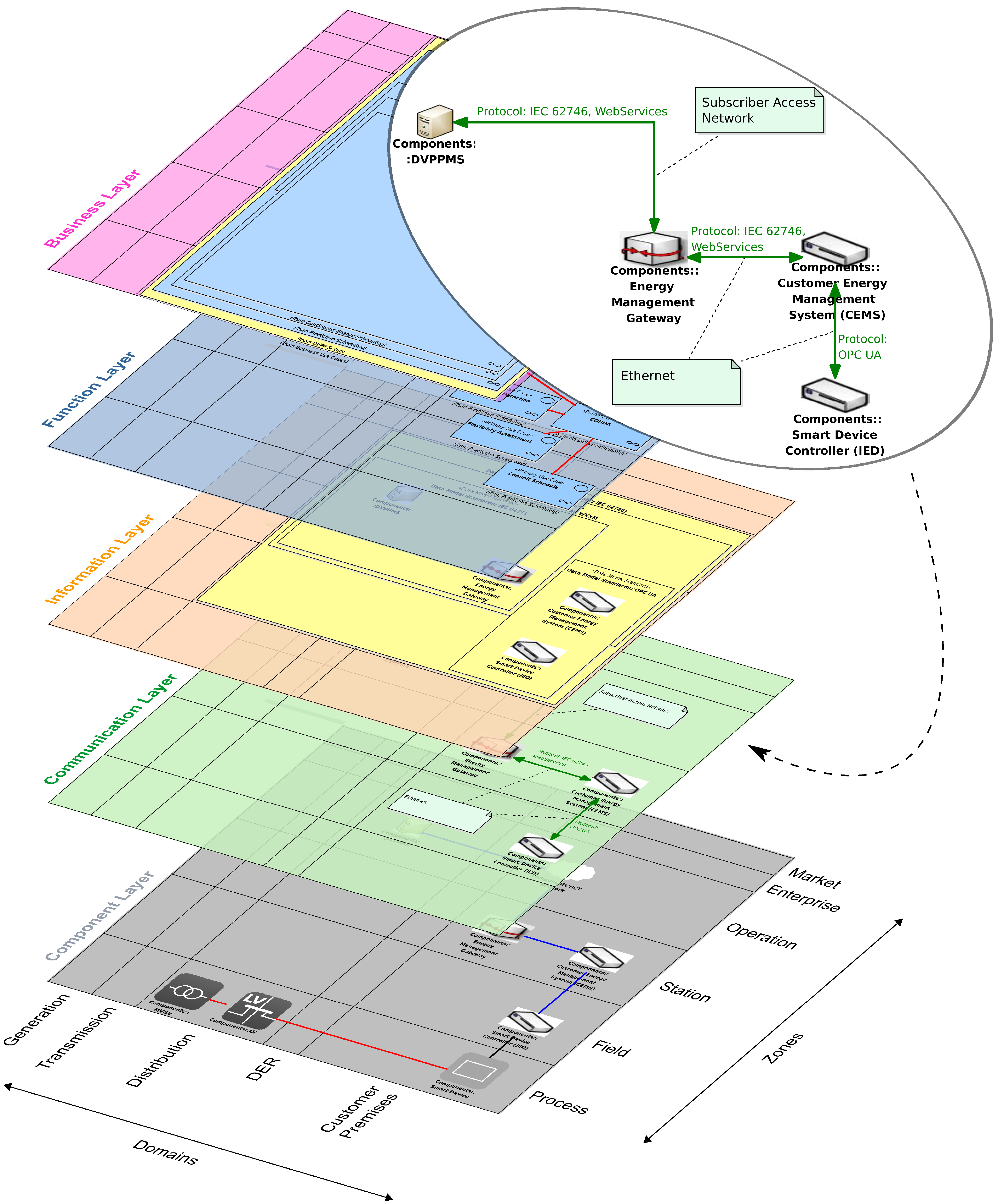}
  \caption{Architecture model of the HLUC \emph{Predictive Scheduling}, showing all interoperability layers of the SGAM framework in a stacked view. For demonstration purposes, the communication layer is shown as an extracted and magnified image without background coloring above the 3D graphic.}
  \label{fig:SGAM}
\end{figure}

 For convenience, we omit the detailed modelling of each layer for our motivating business case here, and provide a visualization of the resulting architecture in Fig.\,\ref{fig:SGAM}.
The figure shows all interoperability layers of the SGAM framework in a stacked view. As the graphical complexity of the figure is rather high, the complete documentation of this design has been made available online.\footnote{\url{http://www.uni-oldenburg.de/en/ui/research/topics/cohda/}} The online version allows exploring the architecture interactively. The user may focus on specific layers or elements, and can access the underlying diagrams such as sequence diagrams, actor mappings etc.
\par In Fig.\,\ref{fig:SGAM}, at the business layer on top, the surrounding business case is depicted with its included HLUCs. The rest of the figure focuses on the HLUC \emph{Predictive Scheduling} only: Obviously, this HLUC spans the from the Process zone to the Operation zone in the information management viewpoint, while being located mainly in the Distribution domain and the Customer Premises domain across the layers. Moreover, the location of the PUCs in the function layer reveals a concentration of the business logic in the Station zone under the Customer Premises domain. Accordingly, suitable components have been identified in the component layer at the bottom of the figure, which serve as hosts for the software applications that realize our business logic. Finally, the information layer and the communication layer present a mapping of international standards to the data flow between the components (see the magnified part in the graphic).
\par In summary, an architecture based on standards could be modelled for the HLUC \emph{Predictive Scheduling} without major incompatibilities. Thus, despite its novel approach of a fully distributed control concept, it can be regarded as compatible to the Smart Grid application domain.
\section{Analysis and Evaluation from a General Perspective}\label{sec:analysis}
As already stated in Sect.\,\ref{sec:conceptualize}, one of the main characteristics of the motivating DVPP business case is the distributed nature of its underlying algorithms. From a general perspective, a distributed algorithm or heuristic defines what, when and with whom to communicate, and what to do with received information, in order to efficiently solve a given problem or task in a distributed manner. Depending on the communication structure, an approach can further be classified as decentralized, hierarchical, distributed or fully distributed \cite{SHNV15}. Moreover, we may distinguish synchronous from asynchronous approaches \cite{Lynch1996}. Especially in the latter, communication irregularities can have a severe impact on the overall progress, because they may change the order of actions in the system that exert influence on each other \cite{HS14b}. So besides performance and efficiency in terms of \eg solution quality, run-time or communication complexity, further criteria are necessary for a proper evaluation in a particular application context. These include \eg robustness analyses and scalability predictions with regard to different problem-specific parameters. In order to facilitate a structured evaluation approach, this section introduces a taxonomy of evaluation criteria, followed by an overview of suitable evaluation methods.
\par Before presenting our taxonomy of evaluation criteria, we have to define a few terms. In compliance with the SGAE process model, we understand a scenario as a specific collection of Smart Grid components, which then constitute the actors that the heuristic under evaluation operates on. These components may be configured using a set of parameters. Then an instance of such a scenario is a parameter assignment for all components within the scenario. Finally, an experiment comprises one or more computational executions of a scenario instance.
\par
\begin{figure}
  \centering
  \includegraphics[width=0.97\textwidth]{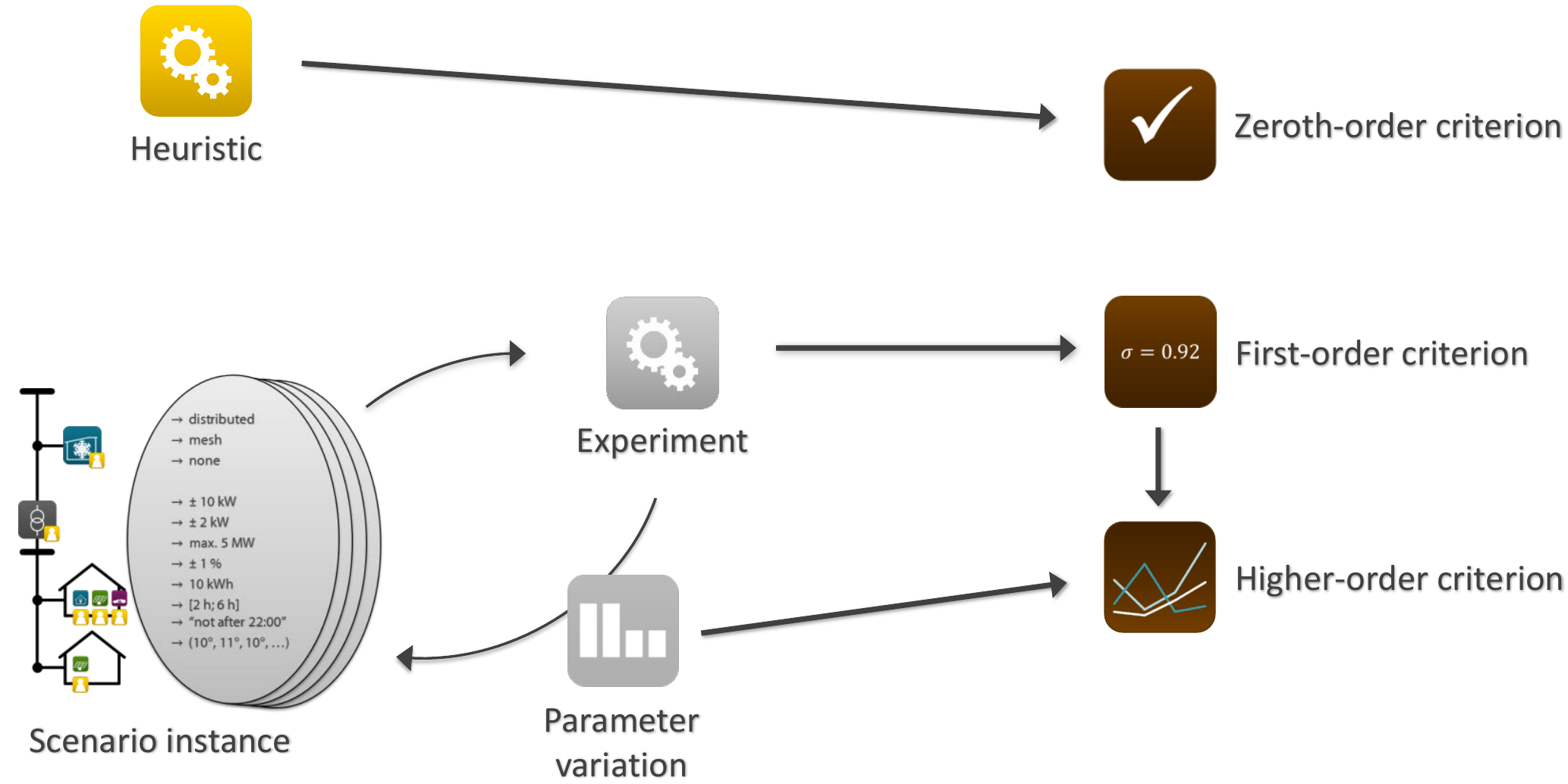}
  \caption{Evaluation criteria types in the context of experimentation.}
  \label{fig:taxonomy}
\end{figure}

 With respect to their dimensionality, we classify evaluation criteria into zeroth-, first- and higher-order criteria, \cf Fig.\,\ref{fig:taxonomy}.
In this context, a zeroth-order criterion yields a basic decision, \ie a yes-no answer, which should generally be independent of any scenario configuration. On the other hand, a first-order criterion provides a scalar quantity, which is usually the outcome of an experiment, \ie the interpretation of experimental data from a scenario instance. For a stochastic model, this experiment may consist of several replications with varying uncontrollable external effects, yielding a distribution instead of a single value.
\par Finally, higher-order criteria allow quantifying effects that occur due to interdependencies between different scenario instances and their first-order criteria, yielding higher-order quantities such as vectors or matrices as output values. For this, series of experiments are necessary, in which one or more dependent scenario parameters are strategically varied. We will now describe these types of criteria in more detail.
\subsection{Zeroth-Order Criteria}\label{sec:zeroth-order}
One of the most basic aspects to consider when dealing with approaches that are targeted at the implementation in critical infrastructures is their \emph{correctness} \cite{Apt1981}, which corresponds to a zeroth-order criterion in our taxonomy. In the SGAE process model, such criteria are examined in the \emph{Analyze} phase. From a general point of view, the following distinctions can be made.
\par First of all, showing correctness involves asserting that if the approach yields a solution, then this solution will satisfy a given specification, \eg it is a valid solution for the given problem (partial correctness). An additional requirement is its termination, \ie asserting that the algorithm terminates within a finite amount of time after it has been started (total correctness). In the field of distributed heuristics, this is also known as guaranteed convergence. Moreover, if this behavior additionally is independent of the system's starting conditions, the heuristic is said to be self-stabilizing \cite{Dolev2000}. With respect to Smart Grid applications, one usually wants to show self-stabilization, as the involved autonomous components might be in arbitrary, unknown states when an optimization process is to be started. Moreover, as the occurrence of faults leads the system into arbitrary states, self-stabilization would support such applications to recover from these faults autonomously.
\subsection{First-Order Criteria}\label{sec:first-order}
First-order criteria quantify properties of an algorithm regarding their behavior during run-time, \ie while executing the algorithm. They usually comprise scalar values which are meaningful only with respect to the specific scenario instance in which they have been collected. For stochastic experiments, \ie scenario instances that include uncontrollable varying parameters, several executions can be performed in order to yield a distribution of values for the first-order criterion of interest rather than an insignificant single value (\cf Sect.\,\ref{sec:evaluation-methodical-aspects}).
\par The probably most evaluated criterion in this category is \emph{performance}. The performance of an approach describes a quantification of its ability to achieve its goal \cite{Talbi2009}. Typically, this is measured in terms of solution quality, \eg a fitness value that is calculated using an objective function. Here it is important to maintain a defined frame of reference, such that the measured value can be interpreted properly. For example, an adequate approach would be to determine the theoretically best and the theoretically worst solution for a given optimization problem as upper and lower bounds, and to normalize the fitness value to the interval that is spanned by these bounds. Apart from such general measurements, Smart Grid specific performance indicators play an important role to assess the performance of a heuristic in this field. A structured categorization of such performance indicators is yet to be defined and will be subject to future work.
\par Besides performance, the \emph{efficiency} of an approach is of interest, which describes the resource requirements of an algorithm or heuristic \cite{Talbi2009}. Regarding centralized approaches, this is usually measured in terms of run-time, \eg the amount of ``steps'' an algorithm takes for a given input, and memory, \eg the amount of storage capacity an algorithm consumes while processing its input. For distributed approaches, determining the efficiency is more complicated: Regarding run-time, we have to distinguish the amount of time until the whole system terminates from the amount of ``steps'' the individual system components will take to reach this state. The former can be measured easily by means of real time, and will be an important information regarding the speed of the system in a specific hardware environment. The latter, however, is a more general measure as it determines the amount of work a system has to carry out. In this regard, a common practice is to count the number of calls to the objective function of the optimization problem, in each distributed component respectively. This way, both the individual work of the components as well as the overall effort can be determined in a hardware-independent manner. Finally, an additional evaluation criterion for distributed systems regarding the efficiency are communication expenses. In our motivating use case for instance, we are focusing on autonomous distributed components, which leads to a message-passing paradigm (in contrast to a shared-memory model, in which multiple components possess a common working memory, \cf \cite{Lynch1996}). Following, both the amount of exchanged messages as well as the size of these messages are significant factors for determining the efficiency of such approaches.
\subsection{Higher-Order Criteria}\label{sec:higher-order}
In this category, first-order criteria are evaluated against varying input parameters, in order to quantify correlation effects, or to perform a sensitivity analysis. In contrast to repeating a single scenario instance due to stochastically, \ie uncontrollably varying input parameters (as described for first-order criteria in the previous section), higher-order criteria are obtained by strategic variation of input parameters, such that the first-order criterion of interest is evaluated with respect to changing scenario instances, thus describing trends or gradients of first-order criteria under varying conditions (see Fig.\,\ref{fig:taxonomy}).
\par A prominent higher-order criterion is the \emph{scalability} of an approach \cite{Barr1995}. Here, the influence of a change in magnitude of input parameters on one or more relevant first-order criteria is determined. For example, given a centralized heuristic for calculating the schedule of energy resources for a future time horizon with respect to \eg demand predictions, one could study the effects of the length of the considered planning horizon on the run-time of the heuristic. An example regarding distributed heuristics is the influence of the amount of autonomous components that are present in the system on communication expenses.
\par Another important higher-order criterion is \emph{robustness} \cite{Barr1995}, which determines the influence of incidental disturbances from the environment on one or more first-order criteria. Such disturbances could be either ``dynamic'' incidents at run-time like \eg varying message delays during the execution of a distributed system, or ``static'' perturbations that determine the sensitivity to changing starting conditions.
\par It is natural that higher-order criteria are rather difficult to analyze as they include lower-order criteria in different magnitudes. On the other hand, they are especially important when targeting critical infrastructures such as the power supply system.
\section{Analyze: Deriving Performance Guaranteees}\label{sec:analyze}
In the SGAE process model, the goal of the \emph{Analyze} phase is to formally prove specific properties of the designed algorithm (\ie the predictive scheduling in our example). This corresponds to the correctness proof of zeroth-order criteria or sometimes first-order criteria for the algorithm under development.
\subsection{Methodical Aspects}\label{sec:analyze-methodical-aspects}
In an analytical approach, evaluation criteria are quantified by mathematical calculus, \ie inspecting the inherent design of an algorithm formally. For this, the semantic of the algorithm has to be described rigorously. An overview in this regard is given in \cite{Francez1992}. A popular example here is the I/O automata formalization \cite{Lynch1996}, which explicitly models the behavior of different components of a system through a standardized interface and thus allows for reasoning about the system's progress as a whole. Based on this, well-known proof techniques like \eg variant functions or convergence stairs can be easily applied \cite{Dolev2000}, as demonstrated below. Another approach would be to employ automatic model checkers. Due to the numerous different semantic descriptions and methods that are available in this field, we refer to \cite{Muller-Olm1999} for an introduction.
\par The methods quoted above are particularly useful for zeroth-order evaluation criteria, \eg for deriving convergence and termination properties in the \emph{Analyze} phase of the SGAE process model. But recently, this has been adapted to first-order criteria as well. For example, in the context of self-organizing systems, \cite{Holzer2009} proposes quantitative definitions of the first-order criteria adaptivity, target orientation, homogeneity and resilience. These are based on an operational semantic in principle, which has been extended by stochastic automatons though. This allows for modeling the system's behavior not only in extreme cases (\ie the best and worst cases as in the evaluation of zeroth-order criteria), but also in the average case, which is crucial for quantifying first-order criteria. The deduced average case behavior, however, directly depends on the chosen distribution functions for the stochastic parts of the model. As a consequence, special care must be taken in order to properly reflect the real behavior of the modeled system when employing such a method. Hence, if adequate distribution functions for a given system cannot be derived easily, an empirical study might be more appropriate in these cases. This approach is described in the following section.
\subsection{Application to the COHDA Algorithm}\label{sec:analyze-application}
The selection of properties to examine during the \emph{Analyze} phase depends on the actual business case, but usually focuses on hard constraints like \eg real-time requirements. For the HLUC \emph{Predictive Scheduling}, an exemplary set of properties comprises:
\begin{itemize}
  \item the \emph{termination} of the scheduling algorithm at a point in time \emph{before} the start of the delivery phase of the power product, and
  \item the termination in a \emph{consistent state}, implying the calculation of a \emph{valid solution} for the given problem, \ie the assignment of a feasible schedule to each member of the DVPP.
\end{itemize}
In the present case, these properties can be proven for COHDA using the \emph{Convergence} Stairs method \cite{Dolev2000}. We will sketch the proof briefly. First, the COHDA algorithm is formally described in the style of the I/O automata framework \cite{Lynch1996}, which allows to reason about a system comprising interacting components. On this basis, a number of predicates (\ie the convergence stairs) are formulated in such a way that each subsequent predicate implies its predecessor, while the last predicate altogether realizes the above defined properties. For COHDA, three predicates are needed: The first predicate regards the production of an initial valid solution after starting the heuristic, and thus covers the initial setup phase. The second predicate then considers the series of calculated solutions until a point in time, in which no more solutions are found, such that a unique final solution is calculated by some component in the system (recall that COHDA is a distributed approach in which solutions are calculated asynchronously by autonomous components in parallel). Finally, the third predicate ensures that this final solution will eventually be communicated to all components in the system, and that the system terminates in the resulting consistent state. By proving that each of these predicates completes in a finite amount of time, the second property (\ie the termination in a consistent state) is derived. For the first property, however, the system architecture from the \emph{Design} phase has to be considered. First, some hardware requirements are imposed on the physical components of the system, such that the first predicate will provably complete in an appropriately short amount of time (\eg by requiring a minimal computation power of the CEMS and a communication backend which must deliver message in at least $x$ seconds, where $x$ depends on the remaining time until the product has to be delivered). This suffices to guarantee that the approach yields at least one valid solution for the problem in the given time.
\par Please note that the actual optimization towards satisfying solutions with regard to product fulfillment is not considered in the formal analysis. This is because the employed algorithm is a heuristic approach and cannot inherently guarantee any solution quality. Hence, such soft constraints are subject to the \emph{Evaluation} phase later on. However, while examining the above termination properties, it was additionally proven that the approach exhibits the \emph{anytime} property in the following sense: Whenever a component calculates and publishes a solution for the scheduling problem, this solution will be better than the previous solution the component has been aware of. From a global point of view, the heuristic thus produces better and better solutions (\ie schedule assignments for the Smart Devices) over time, until no more improvement is possible. In combination with the ability to manually initiate a consistent termination of the process at any desired point in time, this property makes COHDA a highly dependable approach in the context of Smart Grid applications.
\par A detailed description of the full proof will be published in a subsequent paper.
\section{Evaluation}\label{sec:evaluation}
Most first-order criteria and higher-order criteria have to be evaluated by empirical methods. In contrast to formal reasoning based on a rigorous semantic description of an algorithm, empirical methods are based on actually executing the algorithm, \ie the heuristic in the scope of this paper, within a dedicated environment. From monitoring such executions, quantitative data can be recorded, whose dissection and interpretation then leads to the valuation of first- and higher-order criteria.
\subsection{Methodical Aspects}\label{sec:evaluation-methodical-aspects}
Evaluation by empirical methods involves a number of subsequent steps: As a single execution of an algorithm usually does not yield enough information to deduce general conclusions about the behavior of the system in the average case, an adequate experiment design has to be defined in the first step. For first-order criteria, this includes \emph{tactical} decisions, such as the number of repetitions of the executions, in order to level out random effects from uncertain environments or uncontrollable parameters. This will increase the confidence level of the deduced insights later on. For higher-order evaluation criteria, additional \emph{strategic} decisions have to be made, such as defining a strategy for the intentional variation of input parameters in order to analyze the heuristic's behavior under varying conditions, \cf the \emph{Design of Experiments} step in the SGAE process model \cite{Niesse2014}. A comprehensive overview on these topics from the perspective of simulation experiments can be found in \cite{Kleijnen2008}. In the context of heuristics, further care has to be taken regarding the type of scenario instance that is to be solved by a heuristic in a series of experiments \cite{Alba2005}. While parts of this, like \eg the magnitude of input parameters, are usually already covered in the described tactical and strategic decisions, the inherent type of an underlying problem instance might be of interest as well. On the one hand, synthetically crafted problem instances can be used. These do not reflect the targeted application field, but are constructed in such a way that specific properties are present in the problem to solve. For example, ``deceptive'' problem instances \cite{Goldberg:1989:GAS:534133} are useful to analyze whether a given heuristic is able to overcome local optima in the search space. This way, a deep understanding of the observed effects can be gained. On the other hand, application-specific problem instances aim at reflecting the target application of an approach as close as possible, such that the system's behavior can be observed directly in in its presumed environment.
\par In the second step, the experiment is actually carried out. This is usually done by means of simulation. Regarding our focus on heuristic approaches for Smart Grid applications in this contribution, the simulation model then comprises both the heuristic under evaluation and the environment this heuristic is executed in. Following, it is of utmost importance to build the model as realistic as needed, \ie such that all relevant interdependencies between the (simulated) environment and the heuristic are incorporated into the model. For example, if a given distributed heuristic is said to be asynchronous based on message passing between components, possible flaws from the underlying communication technology such as message delays or buffer overflows should be anticipated. The other way around, if the outcome of a heuristic affects \eg the power flow in an electricity grid, and the resulting effects are relevant for the evaluation, the grid must be modeled in such a way that those effects are properly accounted for. Again, \cite{Niesse2014} gives further suggestions regarding this topic. There, besides conceptual considerations, the modular Smart Grid simulation framework \textsf{mosaic} \cite{Schutte2012} is given as a tooling example in the SGAE process model. To permit even more realistic simulations, the framework can be coupled with hardware simulators such as the Smart Energy Simulation and Automation (SESA) lab \cite{Buscher2014}.
\par Finally, in a third step, the preceding executions of the algorithm have to be analyzed with respect to the criteria of interest. Especially for higher-order criteria, specific metrics and suitable statistical methods can then be applied, in order to draw conclusions from the possibly vast amounts of recorded data. Examples for methods and metrics regarding various evaluation criteria can be found in \cite{Barr1995,Kleijnen2008,Alba2005}.
\subsection{Application to the COHDA Algorithm}\label{sec:evaluation-application}
With respect to the business case ``Trading Flexibility'', and the COHDA approach in particular, the results of an exemplary simulation scenario are presented in the following. Please note that this only serves for demonstration purposes as a proof-of-concept and by no means describes a complete empirical evaluation of the approach. For this, numerous simulation studies have already been conducted with respect to the above guidelines in the past. We refer the interested reader to \cite{HS14b,HLS14b,HBS13} for details.
\par
\begin{figure}
  \centering
  \includegraphics[width=\textwidth]{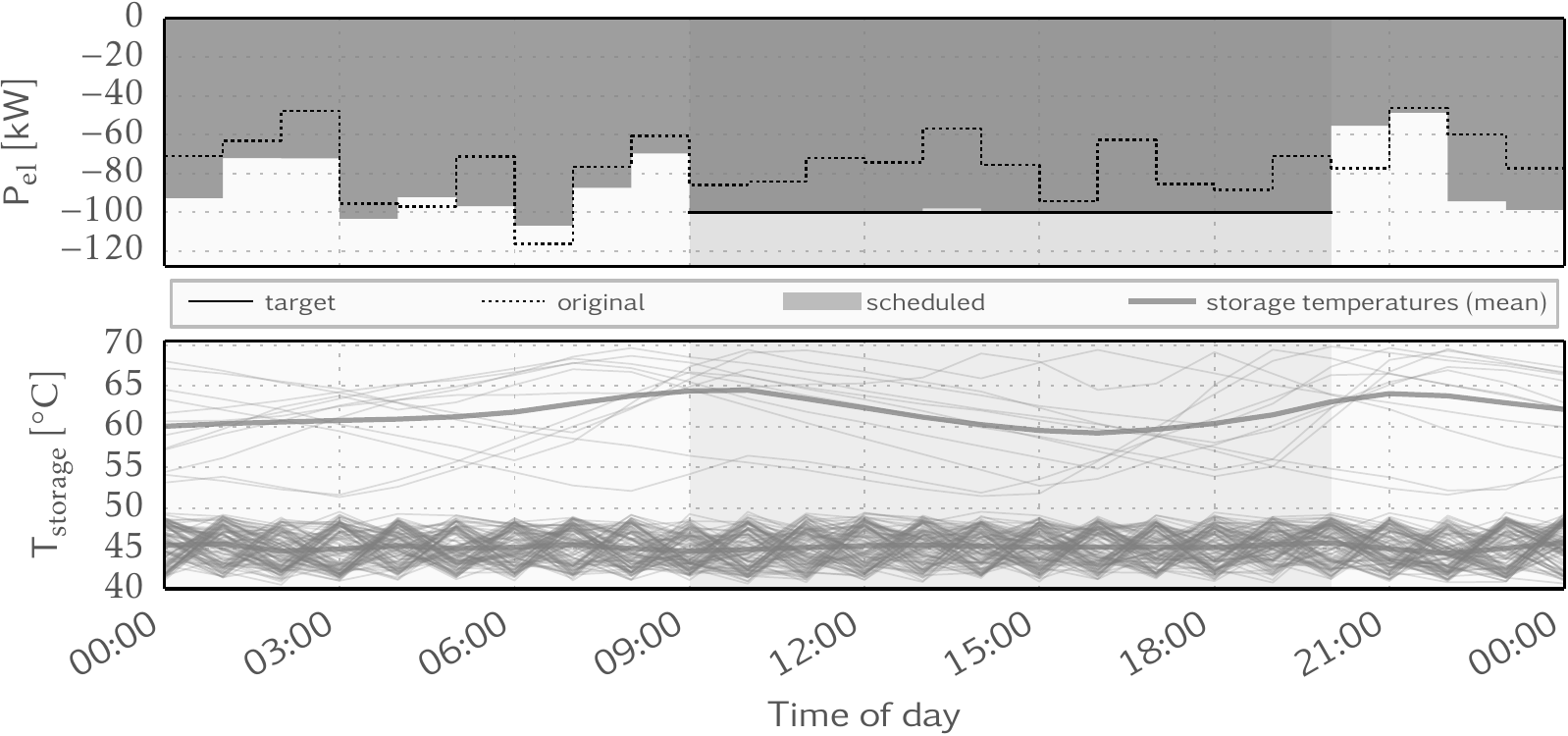}
  \caption{Results of the simulation scenario ``EPEX SPOT Peakload''.}
  \label{fig:Peakload-123-hybrid}
\end{figure}

 In the considered business case, the ESP acts as an intermediate between an energy market and the Flexibility Providers. The European Power Exchange (EPEX SPOT) is an example of a day-ahead spot market for active power in this sense. As a proof-of-concept of the COHDA approach, Fig.\,\ref{fig:Peakload-123-hybrid} shows the scheduling results for a simulated DVPP comprising both flexible loads and controllable small scale generators.
In particular, the DVPP reflects the situation at a typical medium-voltage node in the German power grid. Thus, according to averaged registry data from four large German transmission network operators, it contains 111 geothermal heat pumps with $P_{\mathrm{el}} = -2\,\mathrm{kW}$, 4 combined heat and power plants (CHP) with $P_{\mathrm{el}} = 1\,\mathrm{kW}$ and 8 CHP with $P_{\mathrm{el}} = 5\,\mathrm{kW}$ as Smart Devices (load is depicted as negative power). For these units, we utilized simulation models for the appliances \emph{Stiebel-Eltron WPF 10}, \emph{Vaillant EcoPower 1.0} and \emph{Vaillant EcoPower 4.7}, respectively. As scheduling target, the block product \emph{Peakload} covering the hours 9 to 20 of a trading day was chosen from the list of standardized block products of the EPEX SPOT.\footnote{\url{http://www.epexspot.com/en/product-info/auction}} Reflecting the rather small net power of the DVPP, this target was set to the smallest possible magnitude of $-100\,\mathrm{kW}$ according to the present market rules. In the figure, this target is depicted in the upper chart as a solid line. The relevant time span for this product is illustrated by a slight shading of the background.
\par Hence, the goal for the DVPP was to produce a constant negative amount of $-100\,\mathrm{kW}$ of power during that time span, while being allowed to operate arbitrarily in the remaining time. The simulation covered the full trading day, \ie 24 hours. We excluded stochastic effects like \eg uncontrollable varying thermal demand from this demonstration, thus only a single simulation run is presented. Moreover, because the HLUC \emph{Flexibility Assessment} is not in the scope of this paper, each Smart Device was equipped with a rather simple type of flexibility representation comprising 200 randomly sampled feasible schedules for the respective device (\cf Sect.\,\ref{sec:design} and \cite{HBS13}). But despite this quite limited search space, the heuristic was able to find a schedule assignment that fulfills the target almost perfectly. This is visualized as aggregate power profile of the DVPP by the filled area in the upper plot. For reference, also the uncontrolled profile is shown as dotted line, which was calculated in an additional simulation run without executing the COHDA heuristic. With respect to the presented taxonomy of evaluation criteria, this corresponds to the first-order criterion performance: Interpreted as percentage of the target, the uncontrolled profile would in summary realize $77.35\,\%$ of the target, whereas the profile resulting from the scheduling reaches a coverage of $99.63\,\%$. In the lower plot, the temperature trajectories of the attached hot water tanks are visible. These show that the allowed temperature ranges of the hot water tanks (between $40\,{}^{\circ}\mathrm{C}$ and $50\,{}^{\circ}\mathrm{C}$ for the heat pumps and $50\,{}^{\circ}\mathrm{C}$ and $70\,{}^{\circ}\mathrm{C}$ for the CHPs in our configuration) are not violated by the scheduling actions.
\bibliographystyle{splncs}
\bibliography{main}
\end{document}